\documentclass[aps,prd,showkeys,notitlepage]{revtex4-1}

\usepackage[utf8]{inputenc}
\usepackage{amsmath, amsthm, amssymb, amsbsy,mathrsfs}
\usepackage{comment}
\usepackage{setspace}
\usepackage{graphicx}
\usepackage{hyperref}
\usepackage{pstricks}

\DeclareMathOperator{\Li}{Li}
\newcommand\fverb{\setbox\fverbbox=\hbox\bgroup\verb}
\newcommand\fverbdo{\egroup\medskip\noindent%
            \fbox{\unhbox\fverbbox}\ }
\newcommand\fverbit{\egroup\item[\fbox{\unhbox\fverbbox}]}
\newbox\fverbbox
\newcommand{\be}{\begin{equation}}
\newcommand{\ee}{\end{equation}}

\begin{document}

\title{Coupling the Sorkin-Johnston State to Gravity}

\author{Nicolás Avilán}
\email{ng.avilan47@uniandes.edu.co}
\affiliation{Departamento de Física, Universidad de los Andes,
  Apartado Aéreo 4976, Bogotá, Distrito Capital, Colombia}
\author{Bruno Carneiro da Cunha}
\email{bcunha@df.ufpe.br}
\affiliation{Departamento de Física, Universidade Federal de Pernambuco,
53901-970, Recife, Pernambuco, Brazil}
\author{Andrés F. Reyes-Lega}
\email{anreyes@uniandes.edu.co}
\affiliation{Departamento de Física, Universidad de los Andes,
  Apartado Aéreo 4976, Bogotá, Distrito Capital, Colombia}


\date{\today}   

\begin{abstract}
We consider the dynamics of the Sorkin-Johnston (SJ) state for a massless
scalar field in two dimensions. We conduct a study of the renormalized
stress-tensor by a subtraction procedure, and compare the results with
those of the conformal vacuum, with an important contribution from
correction term. We find a large trace anomaly and compute
backreaction effects to two dimensional (Liouville) gravity. We find a
natural interpretation for the mirror behavior of the SJ state
described in previous works.
\end{abstract}


\maketitle

\section{\label{sec:intro}Introduction}

One of the fundamental problems of Quantum Field Theories in curved
spacetime is the vacuum selection. Unlike maximally symmetric spaces
-- such as the usual Minkowski spacetime, for generic metrics there is
no symmetry argument in order to single out a special, symmetric
state. Questions about the meaning of particle states and decoherence
of vacuum fluctuations, all too important to cosmological studies,
depend upon the solution to this problem to be truly answered.

Recently, Afshordi {\it et al.} \cite{Afshordi:2012jf} made a proposal
for a distinguished ``vacuum'' state of a scalar field for any
globally hyperbolic spacetime. The physical idea behind the selection
was based in the ``causal set program''
\cite{Johnston:2009fr,Sorkin:2011pn}, in which the vacuum state is
associated not with a spacelike slice of a globally hyperbolic
spacetime, but with a causal set of finite size, parametrized by
$\ell$. The existence of a new parameter associated 
with the so-called Sorkin-Johnston (SJ) state codifies beautifully the
physical fact that measurements of quantum states are inherently
finite in space and time, and thus questions related to entanglement
and (de)coherence can be studied from the SJ perspective. In
particular, the entanglement properties of the SJ vacuum could shed
new light to the Reeh-Schlieder theorem.  In its original form the
theorem states that the states constructed by local operators acting
on the usual Poincaré invariant vacuum $|0\rangle$ is dense in the
whole Hilbert space of the theory. It is an interesting question
whether the SJ state satisfy this ``highly entanglement''
property. See \cite{Balachandran:2012pa,Clifton2001} and references
therein for details in this direction.  

From the dynamical perspective, though, the proposal meets some
technical difficulties. The Green's function on this state does not
satisfy the Hadamard condition, and as such fluctuations and couplings
are difficult to tame \cite{Fewster:2012ew}. Backreaction
properties should suffer severely from this fact, and the question on
whether one could define interactions for the SJ state seem now
uncertain. This letter strives to that direction, by studying the
coupling of the SJ state to two dimensional gravity. In two
dimensions, for massless scalar fields, one has the added structure of
a conformal vacuum \cite{BirrellDavies1982}, with which one can easily
compare the properties of the SJ state. We make use of this structure,
as well as the SJ state constructed by \cite{Afshordi:2012ez} to study
the stress-energy tensor and the backreaction to gravity via the
conformal anomaly. This will allow us to conclude that i) the
correction term \cite{Afshordi:2012ez} plays an important role in the
renormalized stress-tensor, almost reducing severely the Casimir
effect associated with the reflecting boundary conditions; ii) there
is strong gravitational backreaction in the SJ state; and iii) once
backreaction is taken into account, the mirror behavior arises
naturally as reflective boundary conditions in the asymptotic induced
geometry.  
  
The letter is organized as follows. In Section \ref{sec:sjprop} we
review the SJ construction for the massless scalar field in two
dimensions. In Section \ref{sec:trenorm} we use point-splitting
techniques to derive a renormalized stress tensor, and in Section
\ref{sec:coupling} we use the result to couple the field to
gravity. We close with concluding remarks in Section
\ref{sec:discussion}. 

\section{\label{sec:sjprop}The Sorkin-Johnston Proposal}

Usual constructions of quantum fields in curved spaces rely heavily on
the symmetries of the spacetime in order to select a maximally
invariant (``vacuum'') state \cite{Wald:2009uh}. Even when it is
applicable, Fock space -- or highest weight -- construction also
suffers from the requirement of non-locality, since mode decomposition
allows for arbitrarily delocalized eigenfunctions of the Laplacian.

These problems motivate the Sorkin-Johnston construction for a state
which can pass as ``vacuum'' for arbitrary (globally hyperbolic)
spacetimes. We will limit the discussion to scalar Klein-Gordon fields
$\hat{\phi}(x)$ . The SJ state $|SJ\rangle$ is a covariant and
uniquely assigned state defined for a real scalar field in a bounded
region of a globally hyperbolic spacetime \cite{Afshordi:2012jf}. Much
like the vacuum state in Poincaré invariant is the highest weight
state for annihillation operators associated with the Klein-Gordon
Hamiltonian, the SJ state is the highest weight state associated with
the mode decomposition of a non-local linear operator:
\begin{equation}
 \hat{\cal S}f(x)=\int_Mi\Delta(x,y)f(y)dV_y,
\label{eq:sj-operator}
\end{equation}
where the kernel $i\Delta(x,y)$ is the Pauli-Jordan function,
\emph{i.e.}, a causal distributional solution to the classical
Klein-Gordon equation. The SJ vacuum is defined in such a way that
upon quantization, the relation $i\Delta(x,y)=\langle
SJ|[\hat{\phi}(x),\hat{\phi}(y)]|SJ \rangle$ holds. The domain of
integration $M$ is a causal set, which we will take to be a causal
diamond. In usual Fock space constructions, the Pauli-Jordan function
is a c-number operator, set by classical dynamics: it is a Green's
function for the Klein-Gordon equation which vanishes outside the
causal cone. The SJ proposal promotes this lack of ambiguity of the
Pauli-Jordan function to a guiding principle behind the definition of
``positive'' and ``negative'' frequencies.

It is proven in \cite{Afshordi:2012jf} that, for any finite region $M$,
the operator $\hat{\cal S}$ is self-adjoint. One can then obtain
eigenmodes $T_k(x)$ such that:
\begin{equation}
 \int_Mi\Delta(x,y)T_k(y)dV_y=\lambda_kT_k(x).
\end{equation}
With this at hand one decomposes the spectrum
into ``positive'' ($\lambda_k>0$) and ``negative'' ($\lambda_k<0$)
frequency modes $T_k^\pm(x)$, and define the SJ Wightman function as:
\begin{equation}
W(x,y):=\langle SJ|\hat \phi (x)\hat \phi (y)|SJ\rangle=
\sum_{k=1}^\infty\lambda _kT_k^+(x)T_k^+(y)^*\quad \lambda_k>0.\label{SJdef}
\end{equation}
Given that the operator is self-adjoint, one can also impose the
reality condition on the eigenmodes $[T_k^-]^*=T_k^+$.

The SJ Wightman function $W(x,y)$ thus defined satisfies these three
conditions \cite{Sorkin:2011pn}:
\begin{itemize}
 \item Comutator: $i\Delta(x,y)=W(x,y)-W^*(x,y)$
 \item Positivity: $\int_{\cal M}dV_x\int_{\cal M}dV_y f^*(x)W(x,y)f(y)\geq0$
 \item Orthogonal supports: $\int_{\cal M}dV_yW(x,y)W^*(y,z)=0$
\end{itemize}
The first two conditions are satisfied by the two-point function of
any state, while the third acts as the ground state condition which
can be interpreted as the requirement that the Wightman function will
be the ``positive frequency part'' of the Pauli-Jordan function,
thought as an (c-number) operator on the Hilbert space of square
integrable functions $L^2({\cal M}, dV)$. The splitting of the Hilbert
space between positive and negative frequency solutions is the
Grassmannian structure at the core of any quantum field theory. The
spirit of defining a quantum field theory from its correlation
functions, rather than the Fock space construction, was anticipated by
the old axiomatic approach, found an explicit representation in the
conformal bootstrap \cite{Belavin:1984vu} and has been incorporated
into recent formulations of quantum field theory. See
\cite{Wald:2009uh} and references therein for a discussion on the
advantages and difficulties. 

The construction thus defines a unique state $|SJ\rangle$ if
$\hat{\cal S}^2$ has a unique square root. This is supposed to hold as
long as $\hat{\cal S}$ is self-adjoint and there are no zero modes. This
condition holds for any bounded region of a globally hyperbolic
four-dimensional spacetime \cite{Afshordi:2012jf}, while for a bounded
region in a two dimensional spacetime the integral operator $\hat{\cal S}$ is
self-adjoint and Hilbert-Schmidt, with finite trace squared
\cite{Afshordi:2012ez}. 

\subsection{SJ state for the massless field in $1+1$ space-time}

We will specialize to a single massless scalar field in $1+1$
dimensions, and chose for $M$ a causal diamond defined by lightcone
coordinates ${\cal M}:\{-\ell\leq u\leq\ell,-\ell\leq v\leq
\ell\}$. The metric and the Klein-Gordon equation is simply:
\begin{equation}
ds^2=-2du\,dv,\quad\quad \partial_u\partial_v\phi(u,v)=0.
\end{equation}
The Pauli-Jordan function in these coordinates is given
by:
\begin{equation}
 i\Delta(u,v;u'v')=-\frac{i}{2}[\theta(u-u')+\theta(v-v')-1]
\end{equation}
such that eigenfunctions for positive eigenvalues of the integral
operator split into two families \cite{Johnston:2010su}:
\begin{eqnarray}
f_k(u,v):=e^{-iku}-e^{-ikv},\qquad &\mbox{ with }&\quad
k=\frac{n\pi}{\ell}, n=1,2,\dots\\
g_k(u,v):=e^{-iku}+e^{-ikv}-2\cos(k\ell),\qquad &\mbox{ with }&\quad
k_n\in {\cal K}=\{k\in {\mathbb R}|\tan(k\ell)=2k\ell\mbox{ and }k>0\}
 \end{eqnarray}
with eigenvalues $\lambda_k=\ell/k$ real. The sum of the squared
eigenvalues is finite ($2\ell^4$), as befitting to an operator of the
Hilbert-Schmidt class. Summing over the two sets of modes,
one finds from \eqref{SJdef} the Wightman
function \footnote{\footnotesize This expression corrects two sign
  typos in \cite{Afshordi:2012ez}.}:
\begin{equation}
\begin{aligned}
W_{SJ}(u,v;u'v') & =\frac1{4\pi}\Big\{-\log\left[1-e^{-\frac{i\pi(u-u')}{2\ell}}
\right]-\log\left[1-e^{-\frac{i\pi(v-v')}{2\ell}}\right]
+\log\left[1+e^{-\frac{i\pi(u-v')}{2\ell}}  \right] \\
&+\log\left[1+e^{-\frac{i\pi(v-u')}{2\ell}}  \right]  \Big\}+\epsilon(u,v;u',v').\\
\end{aligned}
\end{equation}
The $\epsilon$ term is a ``correction term'', stemming from the
fact that the values of $k\in {\cal K}$ are approximately
given by odd multiples of $\pi/2$, $(2n-1)\pi/2$, for sufficiently
large $k$:
\begin{equation}
\epsilon(u,v,u',v')=\sum_{k\in {\cal K}}
\frac{g_k(u,v)g^*_k(u',v')}{k\ell[8-16\cos^2(k\ell)]} 
-\sum_{n=1}^\infty\frac{g^0_n(u,v)(g^0_n)^*(u',v')}{4\pi(2n-1)},
\label{eq:correction-term}
\end{equation}
with $g^0_k = \exp[-(2k-1)\pi i\frac{u}{2\ell}]+\exp[-(2k-1)\pi
  i\frac{v}{2\ell}]$ is the limit of $g_k(u,v)$ as $k\rightarrow (2n-1)\pi/2\ell$.
This term contributes comparatively little for the SJ function, but
will contribute to the stress-energy tensor. It does have a
well-defined coincidence limit $(u,v)\rightarrow (u',v')$ and hence it
can be studied independently of the divergent part. If we disregard
the correction term, the expression for the SJ Wightman function
becomes 
\begin{equation}
W_{SJ,\rm{box}}(u,v;u',v')=\frac{1}{4\pi}\log\left[
\frac{\cos\left(\frac{\pi}{4\ell}(u-v')\right)
\cos\left(\frac{\pi}{4\ell}(v-u')\right)}{
\sin\left(\frac{\pi}{4\ell}(u'-u)\right)
\sin\left(\frac{\pi}{4\ell}(v-v')\right)}
\right],
\label{sj-box}
\end{equation}
exactly the expression for the two point function of a massless scalar field
in a box with reflecting boundaries at $x=\pm\sqrt{2}\ell$ in a
conformal vacuum.

In the large $\ell$ limit, the SJ Wightman function approximates the
usual Minkowski expression, up to an additive constant, for values of
$u$ and $v$ close to the center of the causal diamond. For values
close to the corner, however, there is an interesting ``mirror
behavior'', where the reflecting boundaries become important.  The
consideration of this behavior, as well as a thorough numerical study
of the correction term $\epsilon(u,v,u',v')$, is performed in
\cite{Afshordi:2012ez}.

\section{\label{sec:trenorm}Stress Tensor Renormalization}

In the Poincaré vacuum, the expectation value for the stress tensor is
infinite, and one usually deals with this by assuming normal
ordering. Physically, one could think of the normal ordering as
imposing the constraint that the vacuum should have zero energy and
momentum. When one considers gravity, however, the question of the
vacuum energy is no longer so simple. The stress tensor sources gravity,
and one has to resort to other means to be rid of the infinities. One
such method is by adding a cosmological constant term
\cite{BirrellDavies1982}, which sets some fiduciary, usually
associated to the ``vacuum'' in curved space time, stress tensor
$T_{ab}^{0}$, to zero. The physical stress tensor will then be:
\begin{equation}
T_{ab}^{\rm ren}(x)=T_{ab}(x)- T^0_{ab}(x).
\end{equation}
Of course, once one could change the fiduciary metric or some other
property of the spacetime, $T^0_{ab}$ becomes observable. Such effects
would correspond, for instance, to the Casimir effect or the
dependence of the Cosmological constant to some physical parameter. 
The procedure to define $T^0_{ab}$ is as follows. One starts
with the symmetric Green's function, or the Hadamard
elementary function
\begin{equation}
G^{(1)}(x,x')=\langle \Omega |\{\phi(x),\phi(x')\}|\Omega \rangle,
\end{equation}
for a generic ``vacuum'' state $|\Omega\rangle$, and takes the
derivatives involved in computing the stress energy tensor before
taking the coincidence limit: 
\begin{equation}
T^0_{ab}=\langle T_{ab}(x)\rangle_{\Omega} =\lim_{x'\to
  x}\mathscr{D}_{ab}(x,x')G^{(1)}(x,x'); \quad\quad
\mathscr{D}_{ab}(x,x')=\frac{1}{2}\left[\nabla_a\nabla'_{b}+
  \nabla'_{a}\nabla_b\right]
. \label{Tren}
\end{equation}
In this way one bypasses the ambiguities involved in defining the
product of local operators at coincident points, and take a symmetric
view on the arguments. In our application, the vacuum state is
$|\Omega\rangle = |SJ\rangle$, and  the differential operator
$\mathscr{D}_{ab}$ will have the flat space expression. We
will implement the renormalization by inspecting how \eqref{Tren}
changes with the parameter $\ell$. Because of scale invariance, this
may mean changing the size of the causal diamond or changing the
cosmological constant ``counterterm''. We will delay the discussion
about backreaction to the next Section.

To compute \eqref{Tren} with Hadamard's elementary function
associated with \eqref{sj-box}, we first check how the expression
changes with $\ell$ \cite{BirrellDavies1982}:
\begin{equation}
\frac{\partial}{\partial \ell} T^0_{ab}(x)
=\lim_{x'\to x}\frac{\partial }{\partial
  \ell}\mathscr{D}_{ab}(x,x')G^{(1)}(x,x') \label{TrenD},
\end{equation}
which can be split into the contributions coming from
$W_{SJ,\rm{box}}$ and the correction term, or
$T^0_{ab}=T^{\rm{box}}_{ab}+T^\epsilon_{ab}$. We can check that:
\begin{equation}
\begin{split}
\partial_u\partial_{u'}(W_{SJ,\rm{box}}(u,v;u',v')+W_{SJ,\rm{box}}(u',v';u,v))&=
\partial_{u'}\partial_u(W_{SJ,\rm{box}}(u,v;u',v')+W_{SJ,\rm{box}}(u',v';u,v))\\
& = -\frac{\pi}{32\ell^2\sin^2\left(\frac{\pi(u-u')}{4\ell}\right)}.
\end{split}
\end{equation}
With this expression, the limit $u\rightarrow u'$ diverges, but the
derivative of the stress energy tensor is finite:
\begin{equation}
\frac{\partial}{\partial \ell} T^{\rm{box}}_{uu}(u,v)
=\frac{\pi}{48\ell^3}.
\end{equation}
Thus the finite part of $T^{\rm{box}}_{uu}$ should be:
\begin{equation}
T^{\rm{box}}_{uu}(u,v)=-\frac{\pi}{96\ell^2}.
\end{equation}
It can be checked that $\langle T^{\rm{box}}_{vv}\rangle$ has exactly
the same value. The calculation is entirely similar, but with the
roles of $u$ and $v$ interchanged. This non-zero value is nothing but
the Casimir energy associated with the fact that the SJ Wightman
function $W_{SJ,\rm{box}}$ behaves as if one has reflecting boundary
conditions at $x=\pm\sqrt{2}\ell$.

The coincident limit of the $u,v$ derivative tells
us how the two-point function, and hence the effective action $\langle
(\partial\phi)^2\rangle$ depend on a scale transformation $u\rightarrow
\lambda u$, $v\rightarrow \lambda v$. By general considerations
\cite{BirrellDavies1982,DiFrancesco1997} this is the expectation value
of the trace of the stress-energy tensor. The coincident limit of the
$u,v$ derivative of the Hadamard function is: 
\begin{equation}
\langle T^{\rm{box}}_{uv}(u,v)\rangle = \lim_{(u',v')\to
  (u,v)}\partial_u\partial_vG^{(1)}(u,v;u',v')
=\frac{\pi}{32\ell^2\cos^2\left(\frac{\pi(u-v)}{4\ell}\right)}.
\end{equation}
We note that it depends explicitly on the coordinates. Again, this
should not be a surprise given that the choice of the 
causal diamond explictly breaks Poincaré invariance. It diverges at
the positions of the ``mirrors'' $x=\pm\sqrt{2}\ell$, a fact we will
turn back to in the next Section.

The $\epsilon$ term \eqref{eq:correction-term} has a finite coincidence 
limit and its contribution to the $uu$ and $vv$ component of the
stress-energy tensor can be readily calculated:
\begin{equation}
T^\epsilon_{uu}=T^\epsilon_{vv}=\frac{1}{4\ell^2}\sum_{n=1}^\infty\left[
k_n\frac{4k_n^2+1}{4k_n^2-1}-(2n-1)\frac{\pi}{2}\right]=
\frac{\sigma\pi}{96\ell^2},
\end{equation}
where $k_n$ is the n-th root of $\tan k=2k$. A numerical estimate
gives $\sigma\simeq 0.938535$. One sees that its effect counters the
contribution of the Casimir energy from $WS_{SJ,\rm{box}}$, even
though the correction term is small and slowing changing over the causal diamond
\cite{Afshordi:2012ez}. The reason behind this apparent contradiction
is that the fraction term in the sum above, coming from the
$L^2$ norm of the eigenmodes $g_n$ enhances the energy for small $n$,
making it larger than the box contribution. We will call the total
contribution a Casimir energy. 

The trace correction due to the epsilon term is also expressed as a
sum over the roots of $\tan k=2k$:
\begin{equation}
T^\epsilon_{uv}=\frac{1}{4\ell^2}\sum_{n=1}^\infty\left[
k_n\frac{4k_n^2+1}{4k_n^2-1}\cos\left(k_n\frac{u-v}{\ell}\right)-
(2n-1)\frac{\pi}{2}\cos\left((2n-1)\frac{\pi(u-v)}{2\ell}\right)
\right]. 
\label{eq:tracebox}
\end{equation}
Using that $k_n\simeq (2n-1)\pi/2-1/((2n-1)\pi)+{\cal O}(n^{-3})$, the
sum can be expressed in terms of polylogarithms. Let us consider
the first two terms of the summand: 
\begin{equation}
\frac{u-v}{2\ell}\sin\left((2n-1)\frac{\pi(u-v)}{2\ell}\right) -
\frac{(u-v)^2}{4(2n-1)\pi\ell^2}\cos\left((2n-1)\frac{\pi(u-v)}{2\ell}\right)
+{\cal O}(n^{-2}).
\end{equation}
The potencially dangerous first term gives a sum of delta functions:
\begin{equation}
(T^\epsilon_{uv})^{(0)}=-i\pi\frac{u-v}{8\ell^3}\left[
\delta\left(\frac{u-v}{\ell}\right)-
\delta\left(-\frac{u-v}{\ell}\right)
\right],
\end{equation} 
whose contribution for states constructed from $|SJ\rangle$ with
smooth test functions is zero. The next order term can be readily
computed from the mode sum, using the Taylor expansion of
$\log(1+e^{iz})$: 
\begin{equation}
(T^\epsilon_{uv})^{(1)}=\frac{(u-v)^2}{8\pi\ell^4}\log\tan^2
\left(\frac{\pi(u-v)}{4\ell}\right), 
\label{eq:epsilontrace}
\end{equation}
which also depends on the coordinates. This term also diverges for
$x=\pm\sqrt{2}\ell$, albeit logarithmically, even though the
correction term is supposed to be small. Terms of higher order will
also display this behavior. Using the identities:
\begin{equation}
\begin{gathered}
\sum_{k\;\rm{odd}}\frac{\cos k
  \theta}{k^n}=\frac{1}{2}\left(\Li_n(e^{i\theta})+
  \Li_n(e^{-i\theta})-\Li_n(-e^{i\theta})-\Li_n(-e^{-i\theta})\right) \\
\sum_{k\;\rm{odd}}\frac{\sin k
  \theta}{k^n}=\frac{1}{2i}\left(\Li_n(e^{i\theta})-
  \Li_n(e^{-i\theta})-\Li_n(-e^{i\theta})+\Li_n(-e^{-i\theta})\right),
\end{gathered}
\end{equation}
one can write the generic term as a polynomial times the combination
of polylogarithms $\Li_n(z)$. The singular behavior of the trace stems
from the expansion of the polylogarithm near $z=1$
\cite{Gradshteyn2007}: 
\begin{equation}
\Li_n(z)=-\frac{(z-1)^{n-1}}{(n-1)!}\log(1-z)+f(z)+(1-z)g(z)\log(1-z), 
\end{equation}
with $f(z)$ and $g(z)$ analytic at $z=1$. One should
expect an logarithm divergence at $x=\pm\sqrt{2}\ell$ for all terms in
the expansion, and then the contribution of the correction term to the
trace will be subdominant with respect to \eqref{eq:tracebox}. Its
effect to \eqref{eq:epsilontrace} will be felt near the divergent
points $x=\pm\sqrt{2}\ell$ by changin the $(u-v)^2$ term in front to a
generic polynomial on $(u-v)$. We will disregard the contribution of
these subdominant terms from now on. 
  
With this provision, let us now sum the different contributions. In
terms of coordinates $t$ and $x$, $T_{\mu\nu}(t,x)$ is given by: 
\begin{equation}
\begin{gathered}
\langle T_{tt}(t,x) \rangle
=-\frac{(1-\sigma)\pi}{96\ell^2}+\frac{\pi}{32\ell^2\cos^2\left(\frac{\pi
      x}{2\sqrt{2}\ell}\right)}+\frac{x^2}{4\pi\ell^4}\log\tan^2\left(
  \frac{\pi x}{2\sqrt{2}\ell}\right), \\
\langle T_{xx}(t,x) \rangle
=-\frac{(1-\sigma)\pi}{96\ell^2}-\frac{\pi}{32\ell^2\cos^2\left(\frac{\pi
      x}{2\sqrt{2}\ell}\right)}-\frac{x^2}{4\pi\ell^4}\log\tan^2\left(
  \frac{\pi x}{2\sqrt{2}\ell}\right).
\end{gathered}
\end{equation}
The off-diagonal term $\langle T_{tx}\rangle$ is zero. 

The result can be written in the form:
\begin{equation}
\langle T_{ab}(t,x) \rangle =
-\frac{(1-\sigma)\pi}{96\ell^2}(\eta_{ab}+2u_{a}u_{b})-
 \left(\frac{\pi}{32\ell^2\cos^2\left(\frac{\pi x}{2\sqrt{2}\ell}\right)}
 +\frac{x^2}{4\pi\ell^4}\log\tan^2\left(
  \frac{\pi x}{2\sqrt{2}\ell}\right)\right) \eta_{ab}.
\label{stress-euclidean}
\end{equation}
where $\eta_{ab}$ is the Minkowski metric and $u^a=(\partial/\partial
t)^a$ a constant time-like vector. The result displays not only the
Casimir energy, but also the contribution from the trace anomaly. Its
effect, even for small values of $x$, overwhelms completely the
Casimir term. The divergence means that we can no longer disregard the
coupling to gravity, at least for finite $\ell$. As we take
$\ell\rightarrow\infty$, the contribution vanishes as one should
expect from the Minkowski vacuum.

\section{\label{sec:coupling}Coupling to Gravity}

The result \eqref{stress-euclidean} shows that the expectation value
for the stress energy in the SJ state diverges for finite $\ell$ at
the positions $x=\pm\sqrt{2}\ell$. This
divergence comes from the non-zero value for the expectation value of
the trace of the stress-energy tensor. Classically, the trace should
vanish because of the scale invariance of the massless scalar field in
two dimensions. The choice of SJ state breaks this scale invariance,
and thus we should have a coupling between the expectation value of
the field and the metric. In other words, we should have induced
gravity.

In order to find this metric, let us recall that diffeomorphism
invariance requires that the stress-energy tensor should be conserved
$\nabla^aT_{ab}=0$. But as we can see with a direct calculation:
\begin{equation}
\partial_x\langle T_{xx}(t,x) \rangle=-\frac{\pi^2 \sin\left(\frac{\pi
      x}{2\sqrt{2}\ell}\right)}{32\sqrt{2}\ell^3\cos^3\left(\frac{\pi
      x}{2\sqrt{2}\ell}\right)}-\frac{x}{2\pi\ell^4}\log\tan^2\left(
  \frac{\pi x}{2\sqrt{2}\ell}\right)-\frac{x^2}{2\sqrt{2}\ell^5
\sin\left(\frac{\pi x}{\sqrt{2}\ell}\right)} ,
\end{equation}
with the flat metric expression. The way around it is to impose the
constraint that, in the induced metric $g_{ab}$ we will have
conservation, with $\nabla_a$ the covariant derivative associated with
$g_{ab}$.

Any two dimensional metric is conformally flat, so we can write the
Ansatz for $g_{ab}$:
\begin{equation}
ds^2=\exp({2\varphi})(-dt^2+dx^2),
\end{equation}
where $\varphi(t,x)$ is the Liouville field, and will set the induced
scale. With $T_{ab}$ as in \eqref{stress-euclidean}, the following
equation ensues from $\nabla^aT_{ab}=0$:
\begin{equation}
\begin{gathered}
 \left(T_{xx}-T_{tt}\right)\partial_t\varphi=0, \\
 \left(T_{xx}-T_{tt}\right)\partial_x\varphi=\partial_x
 T_{xx}.
\end{gathered}
\end{equation}
 The first equation above states that $\varphi$ doesn't depend on $t$ as
 expected, while the second can be readily integrated to:
\begin{equation}
\exp(2\varphi) = \frac{\pi}{32\ell^2\cos^2\left(\frac{\pi x}{2\sqrt{2}\ell}\right)}
 +\frac{x^2}{4\pi\ell^4}\log\tan^2\left(
  \frac{\pi x}{2\sqrt{2}\ell}\right).
\end{equation}
One should note that the $\ell\rightarrow\infty$ limit is taken by a
scaling procedure, where the line element $d\hat{s}^2=\ell^2ds^2$
goes to the proper Minkowskian limit when one takes the size of the
box to infinity. The Riemann scalar
$R=-2e^{-2\varphi}\partial_x^2\varphi$ is a complicated function of
$x$, but it asymptotes a constant curvature 
for $x=\pm\sqrt{2}\ell$:
\begin{equation}
 R= - \frac{8\pi}{\ell^2}+\frac{12\pi}{\ell^4}(x\pm\sqrt{2}\ell)^2+\ldots.
\end{equation}
One can see that the induced metric is the consistency condition for
the two dimensional trace anomaly \cite{DiFrancesco1997}:
\begin{equation}
\langle {T^a}_a\rangle =\frac{c}{24\pi}R,
\label{eq:trace-anomaly}
\end{equation}
where $c=1$ for the massless scalar field.

The geometry of the induced metric is asymptotically that of two
dimensional anti-de Sitter space (${\rm AdS_2}$). The patch of the
causal diamond covers a part 
of ${\rm AdS_2}$, see Fig. 1. The geometry has a
conformal boundary at $x=\pm \sqrt{2}\ell$, which is spatial
infinity. Because of the nature of spatial infinity, lightrays take a
finite time, as counted by the observer sitting at $x=0$ to reach
it. Placing reflecting boundary conditions is natural from a broad
perspective, since it allows for an unitary bulk and boundary
dynamics \cite{Avis:1977yn,Wald:1980jn}. It is not by any means the
unique choice  \cite{Ishibashi:2004wx}. The mirror behavior found in
\cite{Afshordi:2012ez} is at any rate deeply tied with the induced
metric. It would not be compatible with any other type of asymptotic
geometry.

\begin{figure}[htb]
    \centering
    \includegraphics[width=0.5\textwidth]{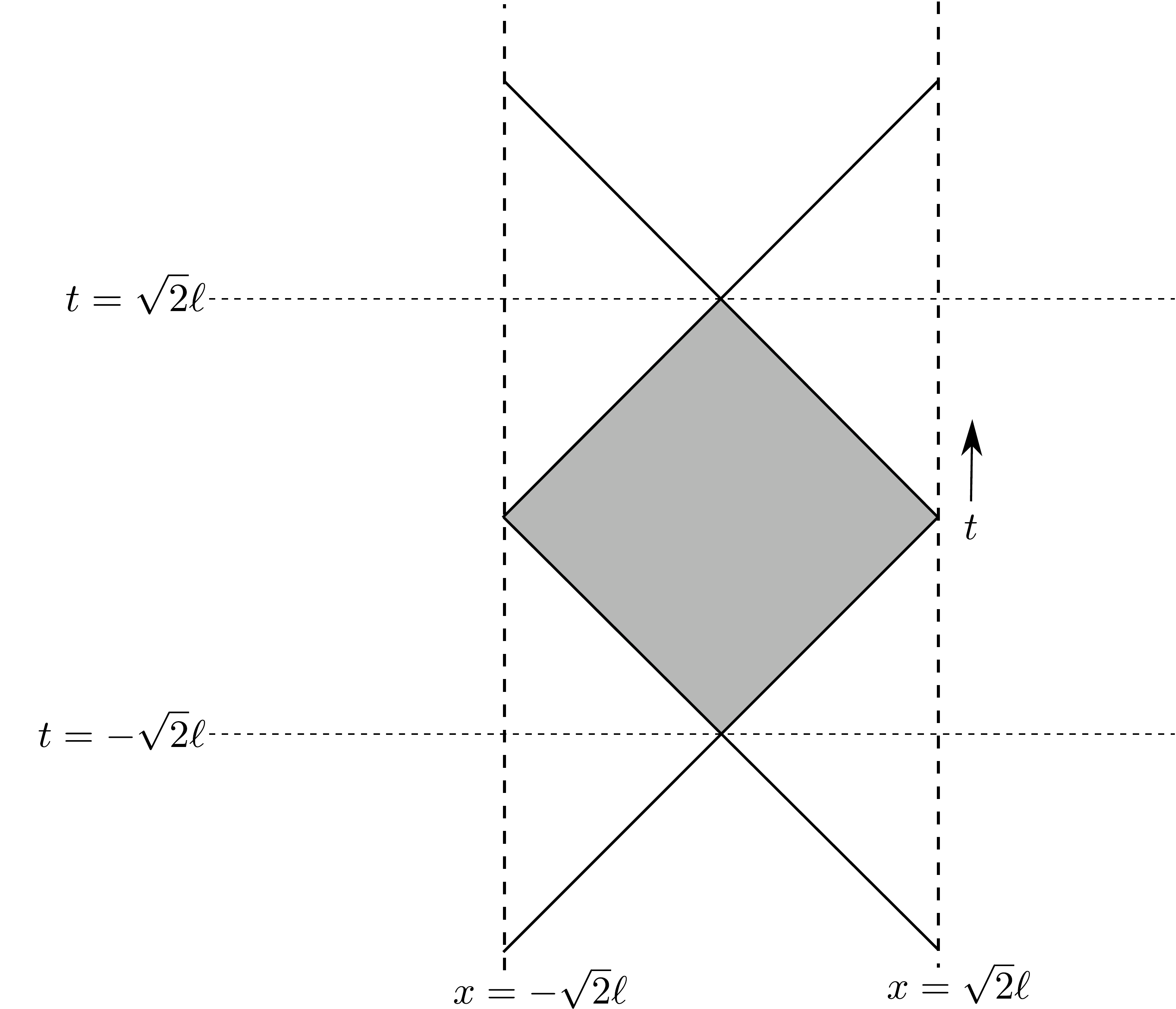}
\label{fig:ads2}
\caption{The causal structure of the induced metric. The causal
  diamond is shaded. The conformal
  boundary sits at $x=\pm\sqrt{2}\ell$ and a light ray takes a finite
  interval of $t$ to cross from one boundary to the other.}
\end{figure}

\section{\label{sec:discussion}Discussion}

In this letter we conducted a study of the stress tensor
renormalization for the Sorkin-Johnston state of a massless scalar
field in two dimensions associated with a causal diamond. We have
found that for the purposes of renormalization of the stress-tensor
energy one can no longer disregard the correction function arising from
finite size effect to the wave number of the modes. The correction
function counteracts the Casimir term, reducing severely its
strength. Moreover, one also finds a trace anomaly term, which  
diverges at the spatial coordinates at the tip of the causal
diamond. Subsequent backreaction shows that the induced metric is that
of a asymptotically anti-de Sitter space with a conformal boundary
sitting at the tip of the causal diamond. This provides another view
on the mirror behavior found previously, corresponding to reflective
boundary conditions for the field in spacial infinity of AdS space. 

From this study one should expect a completely different behavior for
non-zero mass fields: in this case, even classically, the trace is not
expected to be zero, the value is proportional to the mass. It is not
reasonable then to expect the same kind of mirror behavior, as
renormalization effects can be absorbed by a mass redefinition. Also,
questions have been raised about the stability of the linear
perturbations in anti-de Sitter space with respect to backreaction
(see \cite{Friedrich:2014raa} for a recent discussion). Translated to
the SJ construction, these could mean instability of the SJ state
itself. Perhaps this is the reason behind the large contribution of
the correction-term, dominating over the ``box'' Casimir energy. This
should have implications even to the original purpose of defining the
SJ state as a ``vacuum'' which is associated with finite-time
measurements.  

In two dimensions, many problematic aspects of the SJ state do not
appear, which actually makes the whole treatment of this letter
possible. For instance, the treatment in Section \ref{sec:trenorm}
guarantees that the SJ state considered here satisfies the Hadamard
condition. For more generic settings, like the addition of a mass
term, one may not be so lucky. At any rate, the coupling of massive
fields will also have a crucial contribution from the correction term
\eqref{eq:correction-term}, but the analysis can be forced to include
the modification term proposed by \cite{Brum:2013bia}. The
renormalized stress tensor will surely play a big impact on the
understanding of the thermodynamics of the causal diamond, along with
the entanglement entropy. We hope to address these points in future work. 

\section*{Acknowledgements}

We would like to thank Amílcar Queiroz and Fábio Novaes for comments and
suggestions. 

%

\end{document}